\documentclass[chicago]{emulateapj-rtx4}
\usepackage{amsmath}

\makeatletter
\newcommand\an{\ref@jnl{Astron. Nachtr.}}
\newcommand\rma{\ref@jnl{Rev. Mod. Astron.}}
\makeatother

\shorttitle{All-sky Sidereal Tracking With Hexapods}
\shortauthors{P\'al et al.}

\begin{document}

\title{A Hexapod Design For All-Sky Sidereal Tracking}


\author{Andr\'as P\'al\altaffilmark{1,2}}
\author{L\'aszl\'o M\'esz\'aros\altaffilmark{1,2}}
\author{Attila Jask\'o\altaffilmark{1}}
\author{Gy\"orgy Mez\H{o}\altaffilmark{1}}
\author{Gergely Cs\'ep\'any\altaffilmark{3,2,1}}
\author{Kriszti\'an Vida\altaffilmark{1}}
\author{Katalin Ol\'ah\altaffilmark{1}}
\altaffiltext{1}{Konkoly Observatory of MTA Research Centre for Astronomy and Earth Sciences, \\
        Konkoly Thege Mikl\'os \'ut 15-17,
        Budapest H-1121, Hungary}
\altaffiltext{2}{Department of Astronomy, Lor\'and E\"otv\"os University,
                 P\'azm\'any P. stny. 1/A,
                 Budapest H-1117, Hungary }
\altaffiltext{3}{ESO-Garching, Germany,
                D-85748, Karl-Schwarzschild-Str. 2}
\email{apal@szofi.net, lmeszaros@flyseye.net}

\begin{abstract}
In this paper we describe a hexapod-based telescope mount system 
intended to provide sidereal tracking for the Fly's Eye Camera project
-- an upcoming moderate, $21^{\prime\prime}/{\rm pixel}$ resolution all-sky 
survey. By exploiting such a kind of meter-sized telescope mount, 
we get a device which is both capable of compensating for the apparent 
rotation of the celestial sphere and the same design can be used independently 
from the actual geographical location. Our construction is the sole currently 
operating hexapod telescope mount performing dedicated optical imaging survey 
with a sub-arcsecond tracking precision.
\end{abstract}

\keywords{Techniques: photometric -- Instrumentation: miscellaneous}

\section{Introduction}

One of the most challenging part of astronomical instrumentation
is to provide a mount on which the optical system of the telescope
is installed. Basically, such a mount provides both a way to point the
telescope towards a certain celestial position and to compensate
the apparent rotation of the celestial sphere. In modern observational 
astronomy the most commonly used equatorial and alt-azimuth type mounts 
have two-stage serial kinematics. Equatorial mounts are more frequently
used in autonomous or remotely operated sites while the azimuth mechanics
is the standard solution for the ground-based telescopes with
the largest apertures. 

In order to accurately quantify the motion of such a telescope mechanics,
pointing models are constructed. Such mathematical models characterize
how the various mount axes should be actuated to attain the desired 
celestial positions (what are usually defined in first equatorial 
coordinate system: hour angle and declination for the current epoch).
Pointing models exist for equatorial \citep{spillar1993,buie2003},
and alt-azimuth mechanics \citep[see e.g.][]{granzer2012}, as well as
mathematical constructions can be generated for both types of serial
kinematics \citep{pal2015}. 
In addition, accurate pointing models can also be derived if only the attitude 
(orientation) of mount axes are known with respect to the vertical direction
\citep{meszaros2014}. However, this type of pointing feedback can only be 
exploited in the case of equatorial mounts. 

Time-domain astronomy is a novel approach in astronomical 
research that recently became due to the advances in instrumentation.
These advances include the possibility for autonomous operation of 
large telescopes, the employment of mosaic CCD detector arrays
as well as the development of huge computer databases and parallel
data processing systems. The Panoramic Survey Telescope And Rapid Response 
System \citep[Pan-STARRS,][]{kaiser2002} or the (currently under construction)
Large Synoptic Survey Telescope \citep[LSST,][]{ivezic2008} project
are prominent examples of initiatives exploiting these features. 
In these projects it is possible to map the visible sky within a cadence
of a few days and detect variations from this timescale up to several
years of the planned operations. These projects are also noticeable in
their large \'etendue. 

With the increasing technical capabilities it is now possible 
to observe astronomical objects and phenomena on a wide range of time
scales from seconds to years. These kinds 
of research are only possible with surveys that not only run for years, 
but also have frequent temporal sampling. 

The goal of the Fly's Eye project is to provide a continuous, 
$21^{\prime\prime}/{\rm pixel}$ resolution all-sky survey in multiple 
optical passbands by imaging the entire visible sky above the altitude 
of $30^\circ$ with a cadence of a few minutes.
The scientific purpose of the Fly's Eye device is to perform time-domain
observations by covering a time range of nearly six magnitudes with an
imaging cadence of a few minutes up to several years, i.e. the planned length of 
the operations. With this moderate resolution, individual stellar sources 
can be resolved with a sufficient signal-to-noise 
ratio in the magnitude range of $r=7-15$. The scientific goals of such
a simultaneous all-sky survey as Fly's Eye are induced by the continuous 
monitoring of the bright-end ($r\lesssim 15$) regime of astrophysical 
variations and transients. Such signals can either be a small amplitude 
variation of a brighter sources (where the signal-to-noise ratio is limited by
the photon noise or by systematics) or a larger amplitude variation
of a fainter sources (where the signal-to-noise ratio is limited by
the background/readout noise or by confusion).

Regarding to the science domains, this magnitude regime covers 
a/ brighter minor planets in the Solar System, including the homogeneous 
multicolor sampling of the brightest $\sim 10,000$ objects or the flyby 
of near-Earth asteroids; 
b/ stellar astrophysics, including multicolor photometry of pulsating 
and rotating variables, active stars, eclipsing multiple stellar systems, 
transiting extrasolar planets; and 
c/ detection and observations of extragalactic phenomena 
such as brighter supernovae, optical counterparts of GRBs. 

In the Fly's Eye design, the simultaneous surveying of the visible sky 
is performed by a mosaic array of 19 wide-field camera units where each of 
the cameras has a field-of-view of $26^\circ$ in diameter. These cameras are 
assembled on a single, hexapod-based telescope mount. In addition, the total
\'etendue of such a configuration is comparable to that of Pan-STARRS (i.e. 
$\sim 30\,{\rm deg}^2{\rm m}^2$).

Similar initiatives for all-sky surveying include the PASS project 
\citep{deeg2004}. This project has a similar mosaic 
setup (i.e. it uses $15$ cameras to cover the sky above $h\gtrsim 30^\circ$),
but the cameras are fixed to the ground. The 
Evryscope design \citep{law2015}, like its prototype concept
\citep{law2012} features a single-axis setup responsible for sidereal tracking
and employ $24$ wide-field lens. 
Wide-field survey projects that have a comparable field-of-views to the Fly's Eye
were also successful in finding transiting extrasolar planet candidates
\citep[see e.g.][and the references therein]{pepper2007,bieryla2015}.
More specified details of the key scientific projects and design concepts 
can be found in \cite{pal2013}. These goals are in an overlap with the 
goals of the Evryscope project 
\citep[see also the detailed paper of][]{law2015}.

Hexapods are only seldom applied in astronomical instrumentation. 
One of the most prominent usage of such a device is to provide an
alignment and focusing mechanism of secondary mirrors of large 
telescopes \citep[see e.g.][]{geijo2006}. The instrument named
``Hexapod Telescope'' \citep{chini2000} also exploits a similar mechanism
for telescope pointing. This instrument is capable of perform
optical spectroscopy. Probably the most spectacular instrument
driven by a hexapod is the AMIBA mount \citep{koch2009}, a standalone 
radio telescope system. Radio interferometer arrays also 
apply hexapods \citep{huang2011}.

In this paper we describe our design of a small, meter-sized hexapod 
mount dedicated to the all-sky Fly's Eye survey instrument.
In Sec.~\ref{sec:hexapodmount} we detail
how the mount itself is built and controlled. Sec.~\ref{sec:motion} 
describes the basic equations needed for the computations regarding to hexapod
motion control. In Sec.~\ref{sec:tracking}
we describe the computations and algorithms that provide the sidereal 
tracking itself. These computations include the methods needed for automatic 
alignment calibration. In Sec.~\ref{sec:summary} we summarize our results. 

\begin{figure}
\plotone{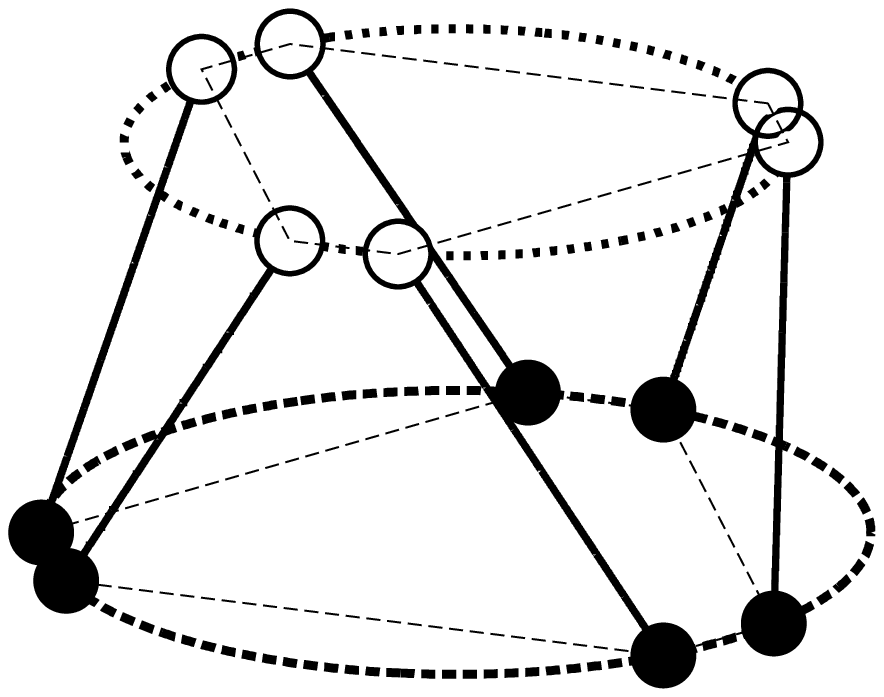}
\plotone{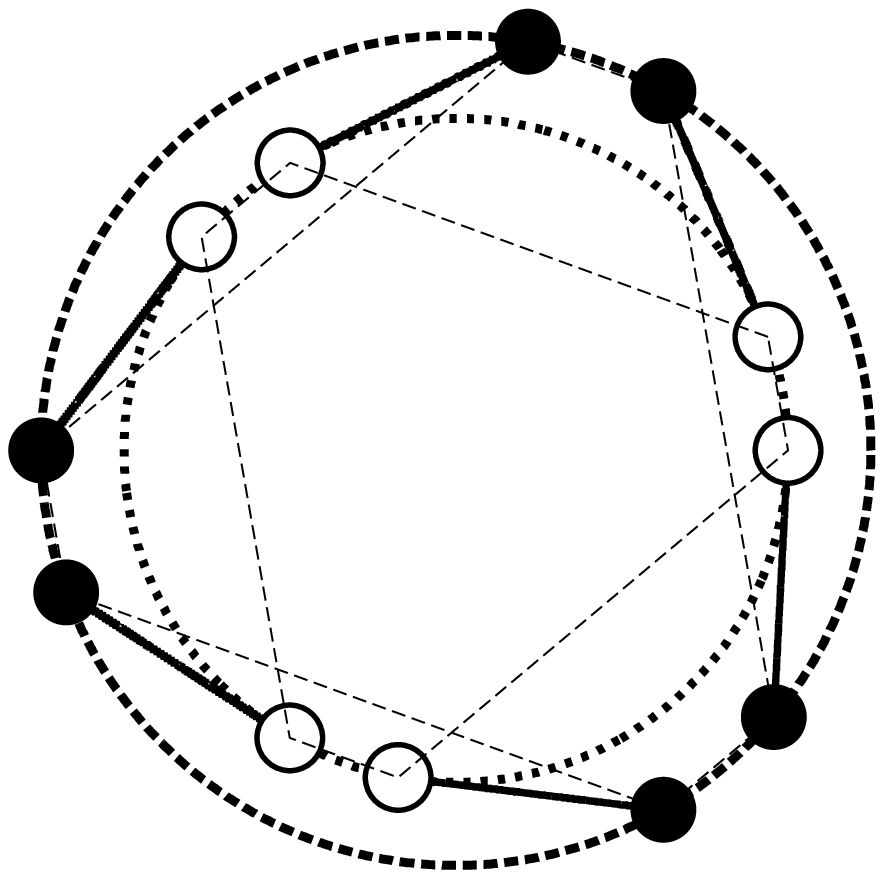}
\caption{A simple sketch depicting the main components of the hexapod.
The upper panel shows a perspective view while the lower panel shows the
similar geometry as viewed from above. The lower/larger dashed circle 
represents the base (that is usually fixed to the ground
which also defines the reference frame for the hexapod kinematics),
the upper/smaller dashed circle represents the payload platform. 
This platform is controlled by the actuation of the six legs, shown as solid 
lines. The legs are connected to the base platform with joints, shown here
as filled (on the base) and empty (on the payload platform) circles. 
The $D_3$ symmetry of the hexapod can clearly be seen on the lower panel,
showing the the two planar hexagonal configuration of the universal joints.
These hexagons are also marked with a thin dashed line.
\label{fig:hexapodsketch}}
\end{figure}

\begin{figure*}
\plottwo{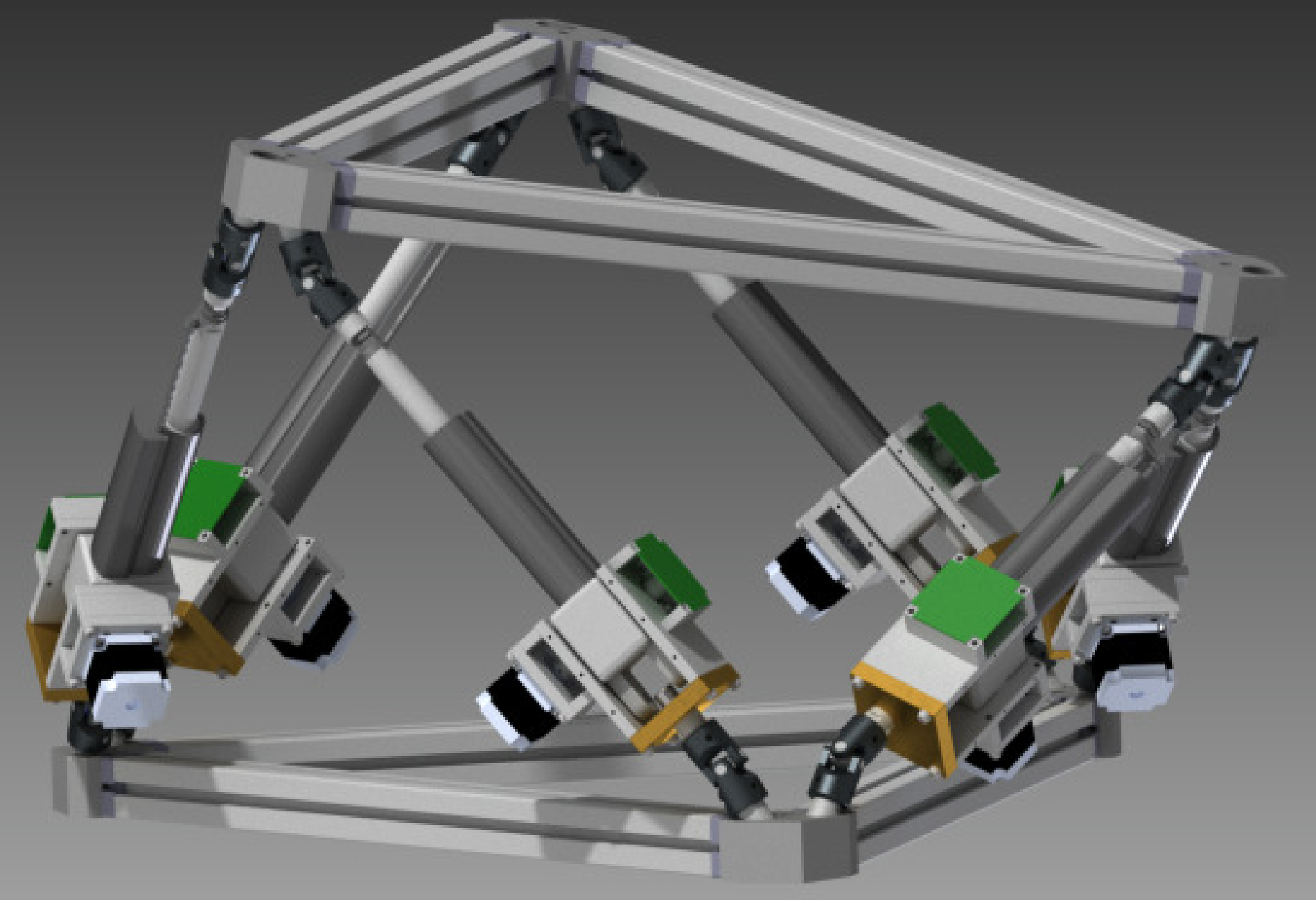}{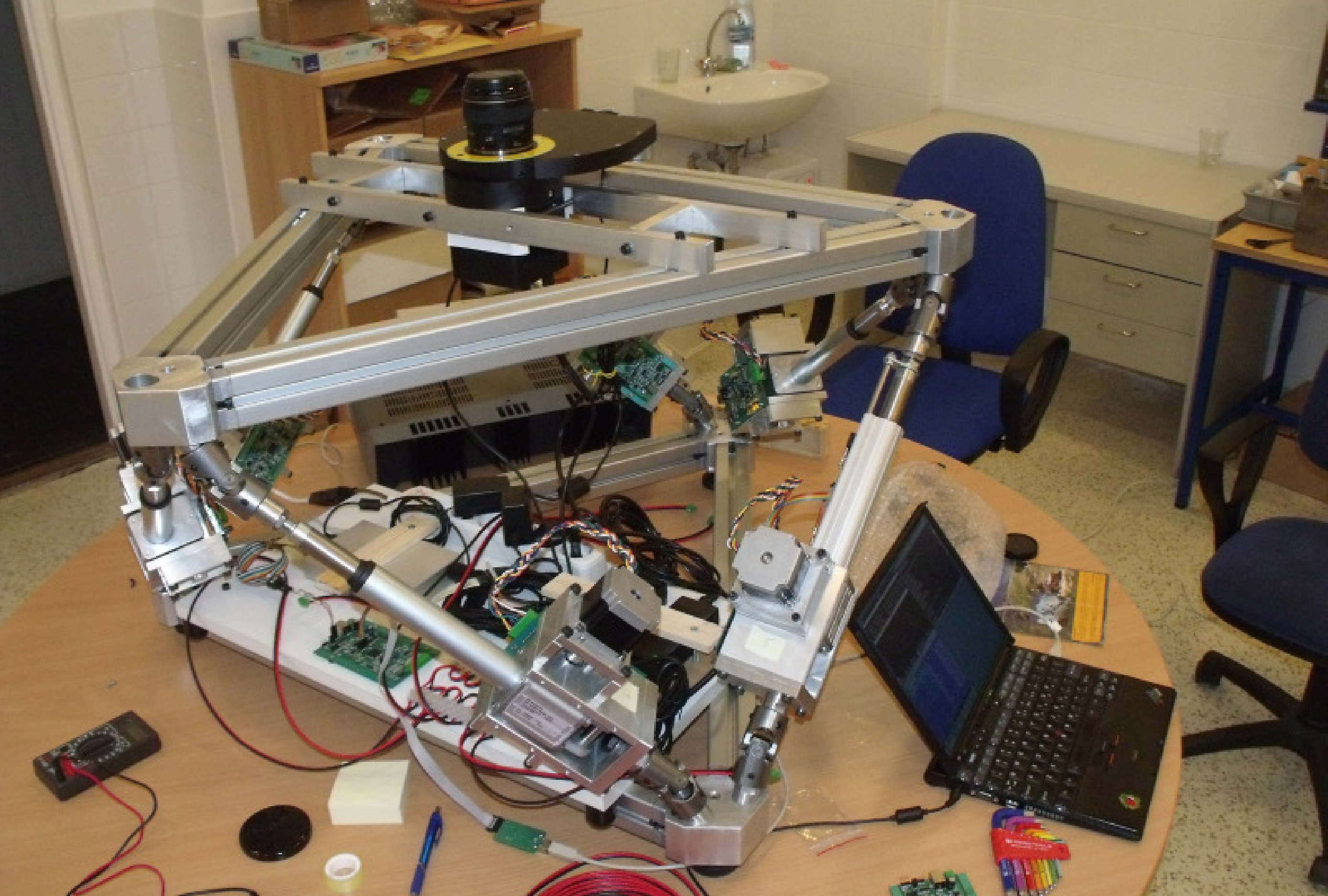}
\caption{Left panel: a computer-aided design (CAD) view of the hexapod 
skeleton, showing the base and payload platforms and the six legs. 
Each leg consists of a high precision
worm gear driven ball screw linear actuator 
(that is driven by a stepper motor) and 
two universal joints. Right panel: the fully assembled hexapod in the lab, 
just prior delivery for first light tests. The hexapod can be 
seen with mounted electronics while additional control
hardware and single-board computer (SBC) is fixed to the base platform. 
The payload is a single imaging camera, 
filter wheel and lens. For simplicity, housekeeping and lens focusing
are managed via spare Serial Peripheral Interface (SPI) and/or 
Inter-Integrated Circuit (${\rm I^2C}$) connectors on the base board.
\label{fig:hexapods}}
\end{figure*}

\section{The hexapod mount}
\label{sec:hexapodmount}

Hexapods, also known as Steward-platforms, employ parallel kinematics 
to provide a complete, 6 degrees of freedom (DoF) spatial motion. 
As its name implies, the motion of a
hexapod is controlled via six legs by altering their lengths.
We note here that in the literature about hexapods, these legs are 
also called \emph{actuator}s, \emph{strut}s or \emph{jack}s. 

These six legs connect the so-called \emph{base} of the hexapod with 
the \emph{payload} platform. By extending or retracting the legs, the 
payload platform will move w.r.t. the base. For instance, simultaneously 
increasing the lengths of all of the legs would move the payload upwards 
while it would not rotate or offset horizontally. 
Throughout this paper, the points which attain the connection between the 
legs and the base/platform are called \emph{control point}s and 
usually realized as a kind of joint (for instance, ball joint or 
universal joint). Fig.~\ref{fig:hexapodsketch} displays a sketch of these 
basic components of the hexapods. 

In the following, we describe the kinematic properties of generic hexapods
in more details. This description is followed by the presentation of our 
hexapod designed for the Fly's Eye project. 

\subsection{Hexapods at a glance}
\label{sec:hexa:glance}

One of the most relevant characteristics of a hexapod is the domain of the 
6 spatial motions in which the payload platform is able to move (w.r.t. 
the hexapod base and/or the reference frame). At the first glance, the 
sizes of these domains (i.e. the total displacement and rotation of 
the platform) are defined by the size of the base/payload platforms, 
the locations of the control points on the base/payload platforms
and the minimum and maximum lengths of the legs. The difference between
the maximum and minimum lengths of the legs is referred also as 
\emph{travel length}.
More precisely, domain sizes depend on the driving mechanism employed 
in the legs as well as the type of joints. 

As a rule of thumb, we can say that the size of the displacement domain for
the payload platform is roughly equal to the travel length while the
size of the rotation domain in radians is roughly equal to the ratio of 
the travel length to the characteristic size of the platforms:
\begin{equation}
\pm S_{\rm platform}\Delta \rho \approx \Delta L\label{eq:rotationdomain}
\end{equation}
Here, $S_{\rm platform}$ denotes the platform size while $\Delta\rho$ 
stands for the rotation domain size and $\Delta L$ is the travel length.

Although any kind of hexapods can be constructed with the topology 
shown in Fig.~\ref{fig:hexapodsketch}, practical designs almost always 
bear various symmetries. Namely, all of the legs are equivalent, by
considering their mechanical design and minimum/maximum strokes while the
arrangement of the 6 control points also shows a triangular ($D_3$) symmetry. 
This arrangement showing the $D_{3}$ symmetry is in fact a planar hexagon with 
two alternating side lengths while all of the internal angles are 120 degrees. 
Hence, this geometry can be quantified by two numbers: the shorter and 
longer side lengths. Depending on the application, the home position of a particular hexapod
is defined either at the state when all of the legs are retracted or 
when the legs are in middle position (i.e. halfway between fully extended and
the fully retracted position).
Considering these types of symmetries, hexapods can be
characterized by $1+2\times 2=5$ geometric parameters at its home position: 
the length of the legs, and the shorter and longer side lengths of
the control point hexagons (both on the base and payload)

More precise qualitative characterization is needed when considering 
hexapods with a particular combination of a leg mechanism and a joint type.
In the following we detail the implications of using a ball screw driven
electromechanical actuator combined with two universal joints on both ends of it.
First of all, one has to ensure that this type of hexapod can unambiguously
be assembled. In the following, we give a short description why this combination
yields a working hexapod. Any solid mechanical part has 6 DoFs
w.r.t. a reference frame. A universal joint has two internal DoFs,
hence 8 DoFs in total. Similarly to a nut on a screw, a ball screw
driven actuator (with a fixed thread) has a single internal DoF, i.e. it has
7 in total. Therefore, considering a disassembled hexapod with a fixed base,
we have $2\times6\times8+6\times7+6=144$ DoFs for all of its parts. 
Assembling two arbitrary parts reduce the total DoFs by $6$. Therefore, 
if we mount the 6 
lower universal joints to the base, the 6 legs to these joints, the 6 upper 
joints to the legs and the platform to the 6 upper joints, we perform
$6\times 4$ assembly operations in total, which reduces the effective
DoFs to $144-6\times 4\times 6=0$. It means that the hexapod can indeed
be assembled in an unambiguous manner. In addition, enabling the ball screw 
threads to rotate in their bearing housing (i.e. to be driven by a motor or 
any external mechanism) allows us to control the hexapod. 

One should note here that hexapods employing ball screw driven actuators
and universal joints suffer from a side effect called 
\emph{screw rotation error} \citep[see e.g.][Appendix B]{koch2009}. This is implied by the 
the fact that the sole internal DoF of the actuator screw thread is
coupled with a rotation of the payload platform. This rotation effectively
changes the length of the leg (i.e. the distance between the control points)
while the screw thread is not driven at all. Since the length of the
leg is usually measured indirectly by encoders placed on the screw thread 
\citep[see][or this paper, Sec.~\ref{sec:hardware}]{koch2009}, this
feedback yields a need for correction. While this side effect can either be
treated as an error, knowing the orientation of the universal joints w.r.t. 
the platforms helps us to properly quantify it. 

\begin{figure*}
\plotone{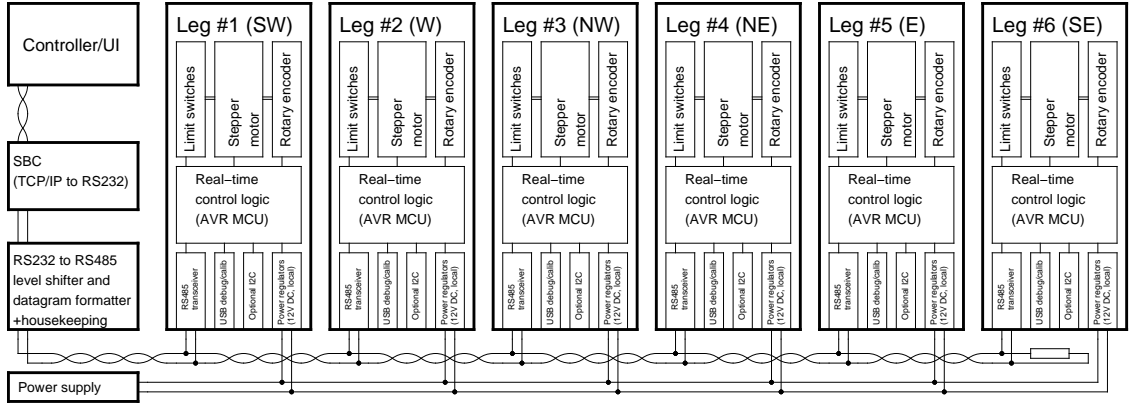}
\caption{Block diagram of the hexapod mount system as it is implemented
in the Fly's Eye project. Each hexapod leg has its own embedded microcontroller
unit (MCU), allowing us to connect the legs directly via higher level
protocols, such as RS485. This setup enables a fully stateless 
operation of the hexapod.\label{fig:blockdiagram}}
\end{figure*}

\subsection{Payload}
\label{sec:payload}

The Fly's Eye design contains 19 camera + filter wheel + lens 
units mounted on a support frame, hence the expected weight of the
payload is approximately $60\,{\rm kgs}$. The CCD detectors have a pixel
size of $9\,{\mu\rm m}\times 9\,{\mu\rm m}$, the imaging resolution is
$4\,{\rm k}\times 4\,{\rm k}$ and we employ lenses with a focal length of
$f=85\,{\rm mm}$ and an f-number of $f/1.2$. This optical and detector
setup yields an effective resolution of $21^{\prime\prime}/{\rm pixel}$. 

The characteristic size of the payload (i.e. the structure that supports
these 19 camera units) is in the range of a meter. From these parameters, 
one can estimate the principal moments of inertia as well as the forces acting 
on each actuator on average. The resolution and repeatability of the 
actuators are implied by the desired sidereal tracking (and/or pointing) 
precision and accuracy. Our requirement for the precision of the sidereal 
tracking is defined as $0.1$ pixels, i.e. $2^{\prime\prime}$. This 
is equivalent to $10$\,microradians 
(note that $1^{\prime\prime}\equiv \pi/648000$ radians).
Therefore the characteristic size of the instrument 
(see also equation~\ref{eq:rotationdomain}) implies
a precision of $\approx 10\,{\mu\rm m}$ on each leg by expecting this
sidereal tracking precision of $0.1$ pixels. 

Since our goal is to perform sidereal tracking during the exposures \emph{and} 
re-adjust the hexapod platform between subsequent images, the 
rotational domain is roughly equivalent (in radians) to the 
exposure time multiplied by Earth's sidereal angular 
rotation frequency. Hence, actuator travel length can be
relatively small compared to the characteristic size of the hexapod. 
A great advantage of the small travel length during operations is that
all actuators are pushed, yielding no backlash in the actuators.

\subsection{Geometry}
\label{sec:geometry}

As we detailed in Sec.~\ref{sec:hexa:glance}, the geometry
of a hexapod bearing similar actuators and a $D_3$ symmetry in both
platforms can fully be characterized by $5$ independent geometric 
parameters at its home position. Due to our intended usage (i.e. performing
sidereal tracking), we declared the hexapod home position when all of the 
actuators are in their middle position. 

The construction for our hexapod (both as a computer-aided design model as well 
as its fully built version) is shown in the panels of Fig.~\ref{fig:hexapods}. 
These images show that our choice for the hexapod contains a similar
base and payload platform. The adjacent sides of the hexagons defined 
by the control points on both platforms have lengths of 
$97.41\,{\rm mm}$ and $742.59\,{\rm mm}$ by design. The lengths of the 
legs were set to $510.00\,{\rm mm}$ at the home position. Our choice
for linear actuators was a model with a net (safe) travel length of 
$\sim 100\,{\rm mm}$. Therefore, the control point distances 
$\ell_i$ ($1\le i \le 6$) can be varied in the domain of 
$460\,{\rm mm}\lesssim \ell_i \lesssim 560\,{\rm mm}$. 
The full rotation and displacement motion domains of our hexapod design
as implied by the geometry described here are summarized in 
Table~\ref{table:motiondomains}. 

\begin{table}
\caption{Motion domains of the Fly's Eye hexapod
in the various rotation and displacement directions. The direction
domains are defined around the home position (where the length of the
legs are $510\,{\rm mm}$). The motion domain limits for the rotations
and displacements are rounded to the nearest tenth of a degree or millimeter,
respectively.}
\label{table:motiondomains}
\begin{center}\begin{tabular}{llrrl}
Direction & Notation & $-$ limit & $+$ limit & unit \\
\hline
Roll 		& $P$		& $-9.5$ & $+9.5$ & ${\rm deg}$	\\
Pitch 		& $\Pi$		& $-9.1$ & $+9.3$ & ${\rm deg}$	\\
Yaw 		& $\Omega$	& $-9.2$ & $+9.2$ & ${\rm deg}$	\\
Left-right 	& $X$		& $-66$	& $+72$ & ${\rm mm}$	\\
Forward-backward& $Y$		& $-75$	& $+75$ & ${\rm mm}$	\\
Up-down		& $Z$		& $-78$	& $+70$ & ${\rm mm}$
\end{tabular}\end{center}
\end{table}


\subsection{Hardware and electronics}
\label{sec:hardware}

As we mentioned in Sec.~\ref{sec:hexa:glance}, our hexapod employs 
legs with two universal joints on both ends while the leg itself is formed 
by a high precision ball screw driven linear actuator. The thread of the
actuator is driven indirectly using a worm gear mechanism with a speed ratio
of $25:1$. The pitch of the ball screw thread is $4\,{\rm mm}$, hence
one revolution in the driven axis is equivalent to $0.16\,{\rm mm}$ travel
length. In order to precisely control the motion, we employed stepper motors
with 200 steps. One full step is therefore corresponds to 
$0.8\,{\rm \mu m}$ of stroke or retract. 

The motors are driven by a high precision controller featuring
sine-cosine microstepping down to $1/16$ steppings as well as a digital
motor current control circuit. A direct feedback from motor motion is 
provided by a full-turn Hall-sensor based encoder with a 12-bit resolution 
(i.e. one full revolution is divided into $4096$ ticks). This resolution
is comparable to the $3200$ microsteps per revolution provided by the 
motor controller. The actuators are also protected by redundant limit switches.

The motor controller as well as the Hall-encoder is driven by an onboard
microcontroller unit (MCU). The firmware of this MCU provides
for each of the legs an embedded control and therefore the legs are 
capable of performing motion in a fully autonomous manner. The firmware 
allows motor positioning with constant speed
as well as ramping with constant acceleration and constant jerk. 
In other words, strokes can be controlled up to a cubic function of the time.
The duration of motion sequences can be commanded in the units of 
$1/256\,{\rm s}$ while the coefficients of the cubic motion are commanded 
in units of motor microsteps. 

The timing resolution of the motion control is $10-15\,{\rm\mu s}$. The
actual resolution depends on the load of the MCU, i.e. in practice,
it depends on the polynomial order of the currently running motion sequence. 
During sidereal tracking, the typical stroke/retract speeds 
are in the range of $\lesssim0.03\,{\rm mm}/{\rm s}$, which is equivalent to 
$\lesssim0.2\,{\rm turn}/{\rm sec}$ in the driven shaft (see above for the 
actual values for thread pitch and gear ratios). Since one turn in the driven 
shaft is divided into $3200$ motor microsteps, sidereal tracking imply a 
$\lesssim600\,{\rm microstep}/{\rm sec}$ stepping frequency. Hence, the 
jitter in the motor speed is less than 1\% by considering microsteps 
(and less than 0.1\% if we consider full motor steps). In addition, the 
corresponding motor control frequency of $60-100\,{\rm kHz}$
is also comparable to the pulse width modulation frequency
of the motor driver circuit (which is set to $\sim 35\,{\rm kHz}$). 

In addition to the aforementioned features, the embedded electronics 
also support a microelectromechanical accelerometer system (MEMS 
accelerometer) on each leg using a similar data acquisition
scheme as it is described in \cite{meszaros2014}. Since the tilt of the 
legs vary as the hexapod moves, these accelerometers provide an independent
means of retrieving the state of the instrument. Due to its complexity,
the mathematics of this hexapod pointing model based on such accelerometers 
are going to be described in a separate paper. 

\begin{figure}
\plotone{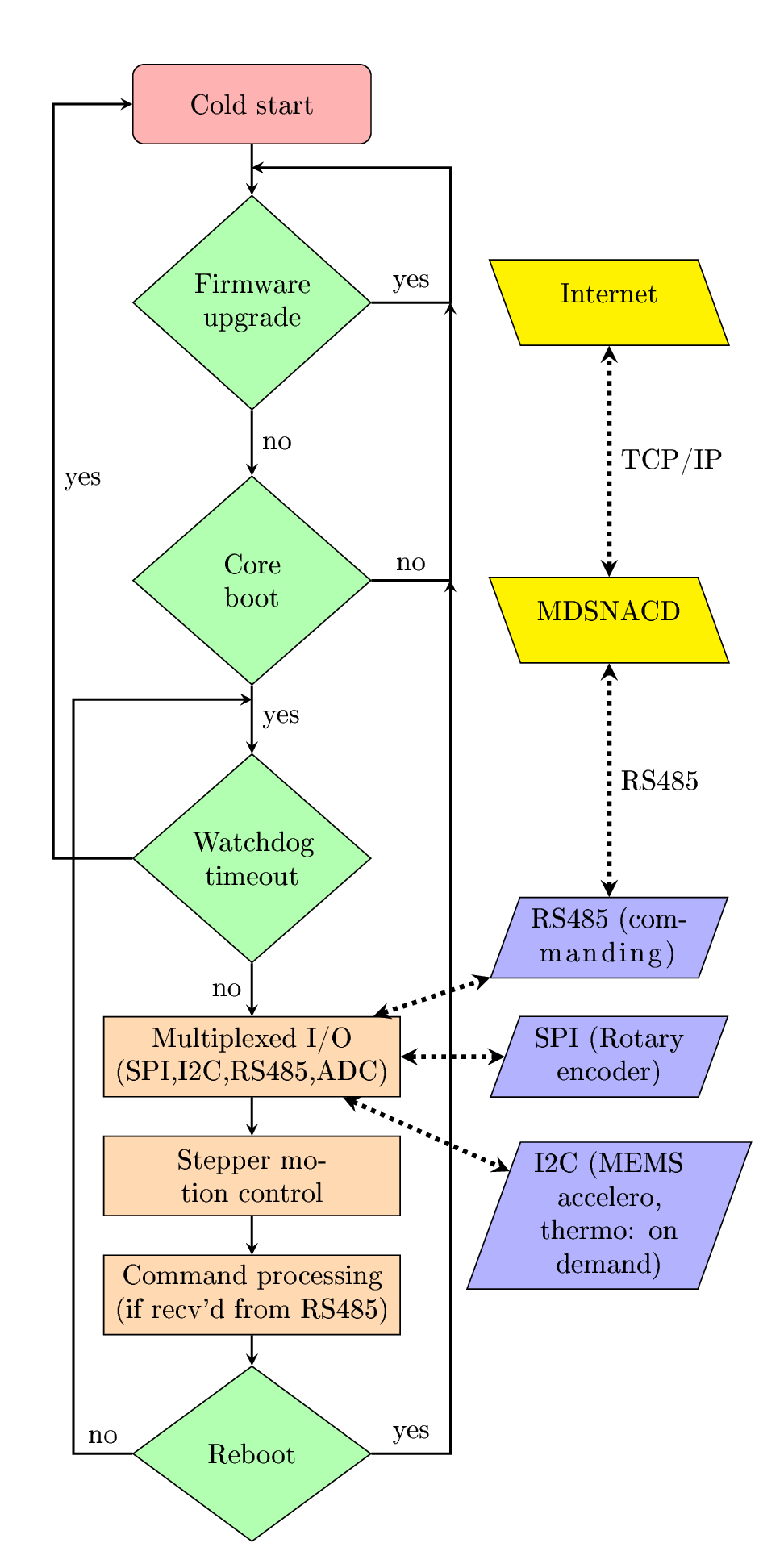}
\caption{A simple flowchart showing the main structure of the MCU firmware
as well as the communication channels between the peripherals (including
the connection with the RS485 bus system). One of the most notable feature
of the structure is the lack of interrupt requests: all of the peripherals
are accessed in a multiplexed form, keeping the duration of a single
processing cycle relatively constant.\label{fig:flowchart}}
\end{figure}

\subsection{Control subsystem}
\label{sec:control}

The microcontrollers are connected to an RS485 bus system in a 
similar fashion used in our former projects \citep{meszaros2014}. 
The RS485 bus system allows the user to upload the coefficients of the 
elementary cubic functions independently for the 6 legs while the 
actual motion can be started simultaneously with a broadcast RS485 command. 
The MCUs are capable of storing a queue with 8 such cubic motion 
sequences in total while new sequences can be uploaded during motion
(if the queue has not been fully loaded). This implementation allows a smooth, 
continuous and fully traceable hexapod motion control. The MCU
firmware also includes features regarding the synchronization of the 
motion (which is needed because of the different primary internal clock 
frequencies). 
The topology of the electronics are displayed in the block diagrams
of Fig.~\ref{fig:blockdiagram}. In Fig.~\ref{fig:flowchart} we show the 
main loops running on the MCU firmware as well as the main communication
channels between peripherals, the RS485 bus system and the higher level
control channels (which is, in practice, a TCP/IP-based protocol, 
implemented in a separate hardware). It should be noted that in the MCU 
core, some of the peripheral communications run constantly (polling and/or
driving the RS485 bus, reading the magnetic encoder, motor step control, etc.) 
while other peripherals are driven only on demand (such as reading the 
thermometer values, accessing the MEMS accelerometer). 
However, the external protocol principle is always master-slave, independently
whether the addressed part is running in a loop or only on demand. 
In other words, this implementation yields a fully autonomous operation but
on the other hand, the hexapod itself (i.e. the six legs, one by one)
must be polled in order to figure out the current status. 

\section{Hexapod motion}
\label{sec:motion}

The hexapod is used primarily for sidereal tracking during exposures in an
all-sky survey instrument. Since a hexapod is
capable of performing arbitrary rotations (within its allowed rotation domain),
it is an ideal instrument to use it for arbitrary geographical locations
and without the need to be perfectly aligned to the principal directions
(i.e. unlike a conventional equatorial mount which needs to be 
aligned properly to the celestial pole).

In order to quantify motions of a hexapod system, first, let us 
investigate the ``small motion'' limit of this construction (with the
geometry described above). Let us quantify the attitude of the platform 
by the orthogonal transformation matrix $\mathbf{O}$, then let us introduce 
the roll, pitch and yaw angles $P$, $\Pi$ and $\Omega$ as
\begin{equation}
\mathbf{O}=\exp
\begin{pmatrix}
0 & -\Omega & \Pi \\
\Omega & 0 & -P \\
-\Pi & P & 0
\end{pmatrix}\label{eq:omatrixexp}
\end{equation}
while the offset vector between the base and platform centers 
is $\Delta=(X,Y,Z)$. Note that in its home position, corresponding to the 
geometry described in Sec.~\ref{sec:geometry}, the roll, pitch and yaw 
angles are $P=\Pi=\Omega=0$, the horizontal displacement offsets are $X=0$ and
$Y=0$ while the distance between the centers of the platform reference points
is $Z=348.35\,{\rm mm}$. 

During routine operations with a hexapod, including the normal
astronomical image acquisition sequences with the all-sky survey instrument,
we have to compute the actuator lengths $\ell_k$ ($k=1,\dots,6$)
as the function of the attitude and offset parameter
vector $(P,\Pi,\Omega,X,Y,Z$ and vice versa. 
The former computation can be deduced with simple vector arithmetics.
Using the notations introduced earlier, the actuator lengths, we got
\begin{equation}
\ell_k=\left\|\Delta+\mathbf{O}\cdot\mathbf{j}^{\rm P}_{k}-\mathbf{j}^{\rm B}_{k}\right\|,\label{eq:actuatorlengths}
\end{equation}
where $\|\cdot\|$ denotes the Euclid norm (length) of the vectors. 
Here the vectors $\mathbf{j}^{\rm P}_{k}$ and $\mathbf{j}^{\rm B}_{k}$
are constants and defines the control point offsets for the actuator $k$ 
w.r.t. the payload platform and hexapod base, respectively. 
Due to the presence of the aforementioned screw rotation error,
the \emph{true} actuator length $\ell_k$ will slightly differ from the 
\emph{commanded} actuator length. 

Both the qualitative and the quantitative behaviour of hexapod motion can well
be analyzed via the derivative matrix
\begin{equation}
\mathbf{L}=\frac{\partial(\ell_1,\ell_2,\ell_3,\ell_4,\ell_5,\ell_6)}{\partial(P,\Pi,\Omega,X,Y,Z)}.\label{eq:l0def}\vspace*{0mm}
\end{equation}
This matrix tells us how the lengths of the hexapod legs alter as one varies 
the payload attitude and/or displacement and vice versa, i.e. how the
attitude and displacement changes when some of the legs are actuated. 
The hexapod do not suffer from any parametric singularity if the 
determinant of $\mathbf{L}$ differs from zero at any arbitrary but allowed
values for $\ell_k$ and/or the $(P,\Pi,\Omega,X,Y,Z)$ parameter vector. 

For the geometry used in the Fly's Eye design (see above, 
in Sec.~\ref{sec:geometry}), the matrix $\mathbf{L}$ 
at the hexapod's home position is going to be
\begin{equation}
\mathbf{L}\approx
{\scriptsize
\begin{pmatrix}
      +0.033   &  +0.287  &    +0.254  &    -0.254  &    -0.287   &   -0.033 \\
      -0.312   &  +0.127  &    +0.185  &    +0.185  &    +0.127   &   -0.312 \\
      -0.307   &  +0.307  &    -0.307  &    +0.307  &    -0.307   &   +0.307 \\
      +0.365   &  -0.730  &    +0.365  &    +0.365  &    -0.730   &   +0.365 \\
      -0.633   &   0      &    +0.633  &    -0.633  &     0       &   +0.633 \\
      +0.683   &  +0.683  &    +0.683  &    +0.683  &    +0.683   &   +0.683 \\
\end{pmatrix}}\label{eq:l0matrix}
\end{equation}
where, for simplicity, the roll, pitch and yaw angles $P$, $\Pi$ and $\Omega$
are measured in milliradians instead of radians. This matrix implies some
of the expected characteristics of the qualitative motion of the hexapod.
For instance, the last row is constant, meaning that a vertical
offset of $+1\,{\rm mm}$ in the payload platform position needs a stroke
of $\approx 0.683\,{\rm mm}$ in all of the legs. However, one should note
that a transformation which essentially swap the $X$ and $Y$ directions
is not a member of the $D_3$ symmetry group. Hence, the lines in the 
above matrix corresponding to the $X$ and $Y$ directions (or similarly, 
roll and pitch rotations) do not show any (implied) correlation between
the respective elements. 

In order to see
that the determinant of  $\mathbf{L}$ differs from zero, let us 
compute the matrix $\mathbf{L}^{\rm T}\mathbf{L}$. Here, $(\cdot)^{\rm T}$
denotes matrix transposition. This computation yields 
\begin{equation}
\mathbf{L}^{\rm T}\mathbf{L}\approx
{\scriptsize
\begin{pmatrix}
 &  +0.295 &  0     &  0     &  0     & +0.279 &  0     \\
 &  0      & +0.295 &  0     & -0.279 &  0     &  0     \\
 &  0      &  0     & +0.565 &  0     &  0     &  0     \\
 &  0      & -0.279 &  0     & +1.600 &  0     &  0     \\
 &  +0.279 &  0     &  0     &  0     & +1.600 &  0     \\
 &  0      &  0     &  0     &  0     &  0     & +2.799 \\
\end{pmatrix}}\label{eq:l0tl0}
\end{equation}
which clearly shows that the determinants of $\mathbf{L}^{\rm T}\mathbf{L}$
and hence that of $\mathbf{L}$ are non-zero. 

The inverse problem related to Eq.~\ref{eq:actuatorlengths} is
the computation of the payload displacement offset and the attitude 
w.r.t. the hexapod base once the actuator lengths are known. This
problem can also be effectively computed in an iterative manner by involving 
the Newton--Raphson algorithm and the derivatives given by $\mathbf{L}$.
The effectiveness of this algorithm is due to the fact
that  $\mathbf{L}^{\rm T}\mathbf{L}$ (and hence $\mathbf{L}$)
varies only slightly in the full domain of 
$460\,{\rm mm}\lesssim \ell_i \lesssim 560\,{\rm mm}$. 
Namely, if we denote the derivative matrix of Eq.~\ref{eq:l0def} 
at the home position (where $\ell_i=510\,{\rm mm}$) by $\mathbf{L}_0$,
then the root mean square of the elements in the matrix
$\mathbf{L}^{\rm T}\mathbf{L}-\mathbf{L}_0^{\rm T}\mathbf{L}_0$ is
around $0.07$, i.e. less by a factor of $\sim 4-40$ than the elements
of the matrix in Eq.~\ref{eq:l0tl0}.

\begin{figure}
\resizebox{!}{40mm}{\includegraphics{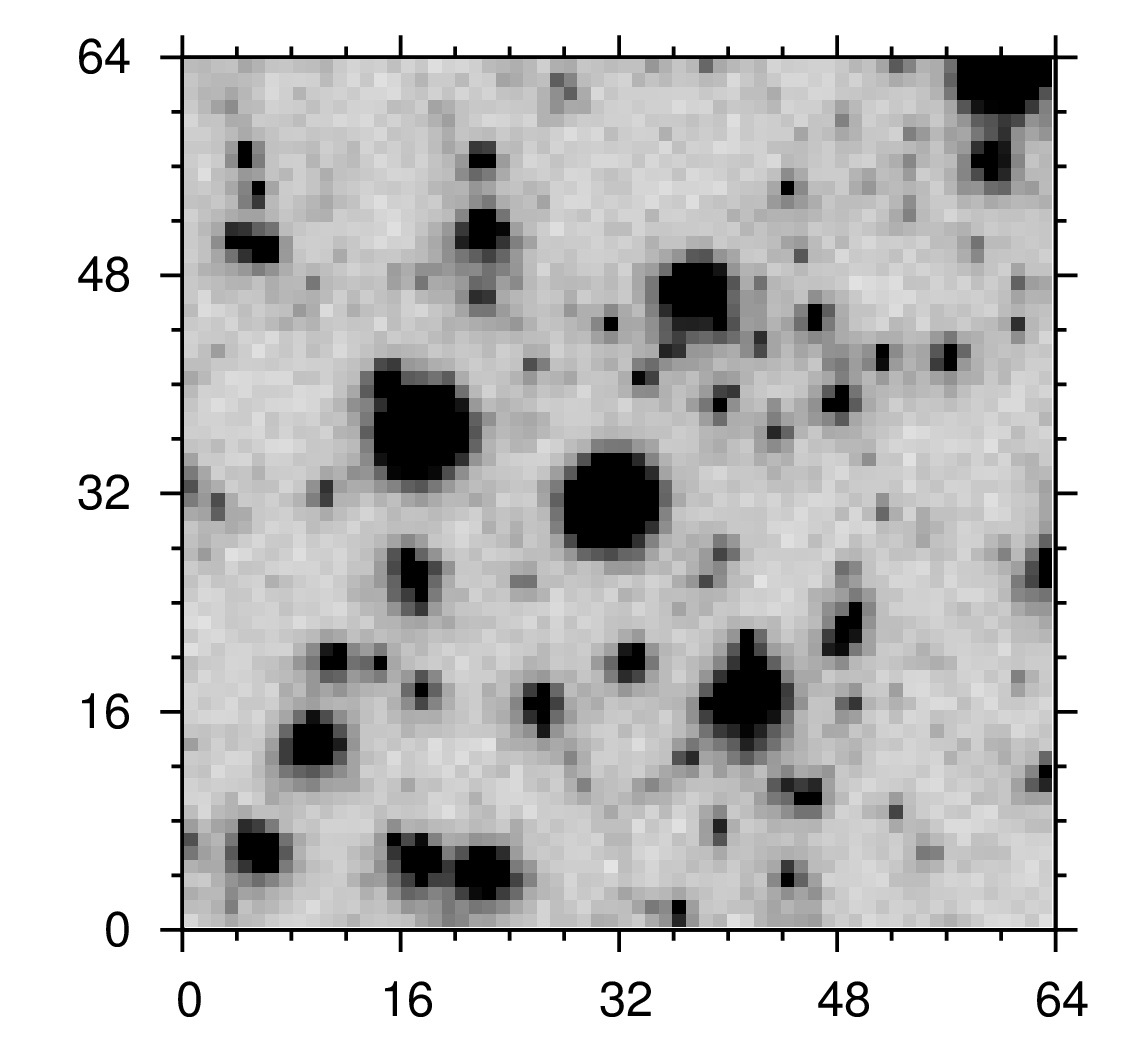}}%
\resizebox{!}{40mm}{\includegraphics{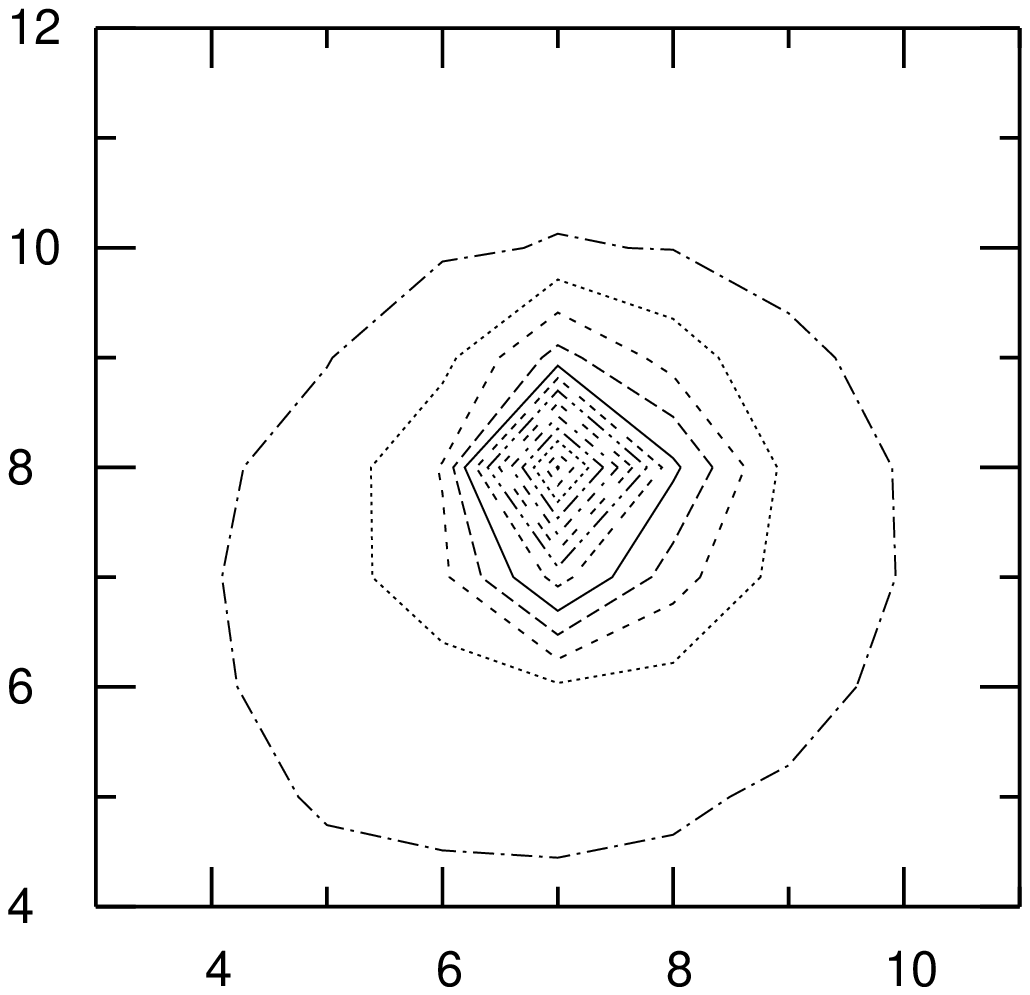}}
\caption{Left panel: a small stamp showing a region of $64\times 64$ pixels,
i.e. approximately $23^\prime\times 23^\prime$ area of the sky. Image
has been acquired during an exposure of $130$ seconds, using an $f/1.8$,
$f=85\,{\rm mm}$ lens and the hexapod for sidereal tracking. 
Right panel: the point-spread function of the stellar profiles shown
on the stamp of the left panel. The PSF is clearly symmetric at its core, the  
wings are due to the aberration of the lens. 
Note that if the sidereal tracking by the hexapod would completely be 
turned off, the lengths of the stellar trails during this $130$ seconds long
exposure would be $\sim 61$ pixels long, i.e. comparable to the size
of the stamp itself.}
\label{fig:stamps}
\end{figure}

\section{Sidereal tracking}
\label{sec:tracking}

Since hexapods are capable of performing arbitrary rotations, this type
of telescope mount can be used without any modifications 
at any geographical location 
as well as the installation procedure does not require precise alignment.
For instance, expecting that the device is installed precisely at the poles,
the apparent rotation of the celestial sphere is compensated by a purely
yaw rotation, i.e. when $P=\Pi=0$ and ${\rm d}\Omega/{\rm d}t = 2\pi/P_{\rm sidereal}$ 
(where $P_{\rm sidereal}$ is the sidereal rotation period of Earth).
Similarly, considering a device installed at the equator, the celestial sphere
would perform a purely roll rotation, i.e. ${\rm d}P/{\rm d}t=2\pi/P_{\rm sidereal}$ while
the two other angles are zero throughout the motion. If the hexapod
is installed at temperate latitudes, then the motion would be a combination
of roll and yaw rotations. 
If the hexapod is not precisely aligned, corrections with small respective 
angular speeds are needed in all of the three principal directions. 

As a foreword, we note here that the algorithm detailed in this section 
works if the conditions of the presented linear treatment meets with the 
accuracy of the initial alignment of the hexapod. In practice, our experience
was that the employment of standard tools (e.g. classic bubble leveler, 
digital compass based on MEMS magnetometers embedded in a smartphone, etc.)
yields a precision comparable to or less than one degree. This accuracy
is sufficient to apply our linear algorithm and yield the sub-arcsecond
tracking precision. However, we should also note that equations
(\ref{eq:omatrixexp}) and (\ref{eq:actuatorlengths}) work on the full hexapod
motion domain (see also Table~\ref{table:motiondomains}). Therefore, 
even if the misalignment is in the range of several degrees, the
similar tracking precision can be achieved by an 
algorithm which composes the matrix exponent of 
Eq.~(\ref{eq:omatrixexp}) with the linear terms described in the following. 

In the following subsections, we present a simple algorithm how a hexapod 
can be calibrated in order to obtain adjustment parameters needed for proper
sidereal tracking. This description is followed by the presentation
of a series of tests validating our algorithm and, in general,
the usability of a hexapod-based mount for all-sky tracking. 

\subsection{Self-calibration}
\label{sec:selfcalibration}

One of the most straightforward ways to calibrate the hexapod motion
parameters is to involve the sky itself as a reference frame. First, let us assume
that the hexapod, located at the geographical latitude $\varphi_0$,
is roughly oriented and aligned to the cardinal directions
as well as the base is nearly horizontal. In this case, the speed of the
leg $k$ at a linear approximation is need to be
\begin{equation}
\frac{{\rm d}\ell_k}{{\rm d}t} = n_{\rm sidereal}\left(\cos\varphi_0\frac{\partial\ell_k}{\partial P}+\sin\varphi_0\frac{\partial\ell_k}{\partial\Omega}\right),\label{eq:speedideal}
\end{equation}
where $n_{\rm sidereal}= 2\pi/P_{\rm sidereal}$, the sidereal angular 
frequency of Earth and the partial derivatives 
$\partial\ell_k/\partial P$ and $\partial\ell_k/\partial\Omega$ can be 
taken from Eqs.~(\ref{eq:l0def}) and (\ref{eq:l0matrix}).

In our construction, the hexapod can directly be commanded to move the 
legs according to the speeds defined by Eq.~(\ref{eq:speedideal}), 
see Sec.~\ref{sec:control} above. 
Since the real placement of the hexapod is not perfect, the
tracking which employs the speeds defined by this equation would also not
be perfect. If one obtains a series of images, the drift as well as the 
field rotation can easily be measured by finding the plate solutions
for these images. We note here that the hexapod also acts as a field rotator
device, hence plate solution should also include information about the
field rotation. Here, these ``series'' should mean at least two images,
however, the longer the series are, the more accurate the determination
of the plate drift and the apparent field rotation. Both the plate drift
as well as the field rotation has a unit of radians per second, i.e.
the actual apparent drift and/or field rotation between the subsequent images 
first must be converted to radians as well as divided by the image cadence.
If the series contains more than two images, these parameters are derived
by simple linear least squares regression. Let us denote these three
angular speeds by $\omega_1$, $\omega_2$ and $\omega_3$. 
For instance, if all of these three parameters are obtained using a pair
of images with a time separation of $\Delta T$, these angular 
speeds are computed as 
\begin{eqnarray}
\omega_1 & = & \frac{S\Delta x}{\Delta T}, \\
\omega_2 & = & \frac{S\Delta y}{\Delta T}, \\
\omega_3 & = & \frac{\Delta\varphi}{\Delta T},
\end{eqnarray}
where the (linear part of the) differential plate transformation 
$(x,y)\to(x',y')$ between 
the pairs is described as 
\begin{equation}
\binom{x'}{y'}=\binom{\Delta x}{\Delta y}+
\begin{pmatrix}
\cos\Delta\varphi & -\sin\Delta\varphi \\
\sin\Delta\varphi & \cos\Delta\varphi 
\end{pmatrix}
\cdot
\binom{x-x_c}{y-y_c}
\end{equation}
$S$ is the plate scale (in the units of arcseconds per pixel), 
and $x_c$ and $y_c$ are the field centroid coordinates (usually the half
of the image size dimensions). If tracking is nearly fine, then 
these values of $\Delta x$, $\Delta y$ (field center offsets) and 
$\Delta\varphi$ (field rotation) are usually
small -- some pixels or some tens of pixels, or equivalently some arcminutes --,
so the linear differential astrometric transformation is still reasonable. 

In the next step, we perturb Eq.~(\ref{eq:speedideal}) with adequately 
large offsets $A_1$, $A_2$ and $A_3$ as 
\begin{eqnarray}
\frac{{\rm d}\ell_k^{(A)}}{{\rm d}t} & = &  n_{\rm sidereal}\left[(\cos\varphi_0+A_1)\frac{\partial\ell_k}{\partial P}+\right. \nonumber \\
& & +\left.(\sin\varphi_0+A_2)\frac{\partial\ell_k}{\partial\Omega}+A_3\frac{\partial\ell_k}{\partial\Pi}\right],\label{eq:speedperturbed}
\end{eqnarray}
In a similar fashion as it was described above, the corresponding 
$\omega_m^{(A)}$ drift and field rotation speeds can also be retrieved by 
exactly the same procedure. It can be considered that the apparent
drift and field rotation vanish if the parameters $A_1$, $A_2$ and $A_3$
are set to satisfy the equation
\begin{equation}
\omega_m + \sum\limits_{k=1}^3A_k\frac{\partial\omega_m^{(A)}}{\partial A_k}=0\label{eq:omegaequation}
\end{equation}
for all $m=1$, $2$ and $3$. Using real measurements on the stellar field
seen by a camera mounted on the hexapod, the partial derivative 
$\partial\omega_m^{(A)}/\partial A_k$ can be measured numerically by
involving the formerly obtained series of test images. Namely,
this derivative can be approximated as
\begin{equation}
\frac{\partial\omega_m^{(A)}}{\partial A_k}\approx \frac{\omega_m^{(A_k)}-\omega_m}{A_k}.\label{eq:diffapproxlefthanded}
\end{equation}
Here $\omega_m^{(A_k)}$ corresponds to a drift and field rotation speed 
derived from an image series when the hexapod was moved according to
Eq.~(\ref{eq:speedperturbed}) and $A_k$ is non-zero while the other two 
$A_x$ ($x\ne k$) values are zero. All in all, four image series should be 
taken in total, corresponding to the speeds $\omega_m$, $\omega_m^{(A_1)}$, 
$\omega_m^{(A_2)}$ and $\omega_m^{(A_3)}$. We note here that the 
derivatives of Eq.~(\ref{eq:diffapproxlefthanded}) can even better be approximated
using symmetric first derivatives instead of this one-sided numerical
derivative. Namely, this approximation is
\begin{equation}
\frac{\partial\omega_m^{(A)}}{\partial A_k}\approx \frac{\omega_m^{(+A_k)}-\omega_m^{(-A_k)}}{2A_k}.\label{eq:diffapproxsymmetric}
\end{equation}
In order to obtain the derivatives in this way, six image sequences
are needed to be taken instead of four. 

One of the most important property of the above procedure is that it does not
need any kind of assumption from the underlying hexapod. Namely, the 
procedure also compensates for the screw rotation error and even we do not
have to precisely know the hexapod geometry as well. In addition,
we need to use at least three hexapod legs during the calibration procedure,
i.e. the number of involved hexapod legs can be decreased until
the corresponding linear combination of the $\mathbf{L}$ matrix components
have a rank of 3 (otherwise, Eq.~\ref{eq:omegaequation} would also be singular).
This fact has an important consequence which can be highly relevant 
during the operation of an autonomous and/or remote observatory such as 
the Fly's Eye device: the sidereal tracking can be attained even if one, two
or three of the hexapod legs are stuck (due to, for instance,
mechanical or electronic failures). 

Moreover, in the case of astronomical imaging, the need for only 3 DoFs out 
of the 6 allows us to vary the displacement of the hexapod payload
independently from its rotation. This can be used to program the hexapod
at different displacement offsets in order to avoid the wearing of the
actuator's ball screws at certain positions. 

\subsection{Absolute calibration}
\label{sec:absoltue}

The above presented algorithm can be extended to attain an absolute positioning
of the hexapod w.r.t. the celestial reference frame. In order to 
accomplish this step, one has to obtain the apparent coordinates of the
stellar field seen by the cameras. After deriving the astrometric solution
of the field in J2000 system, the equatorial coordinates (right ascension and 
declination) are needed to be computed after correcting for precession, 
nutation, aberration and refraction using the standard 
procedures \citep{meeus1998,wallace2008}. These coordinates are then needed
to be compared with the local sidereal time and the geographical latitude.
In a similar fashion, the field rotation with respect to the J2000 
reference frame should also be converted to the field rotation w.r.t.
the apparent north direction. 

It should be noted that the conversion between the differences in the
celestial coordinates or field rotations and the respective roll, pitch
and yaw offsets must be done with caution due to the 
singularity in the parameterization of the the celestial coordinate systems.
Namely, let us define the matrix function $\mathbf{A}(\alpha,\delta,\rho)$ 
as 
\begin{eqnarray}
\mathbf{A}(\alpha,\delta,\rho) & = & 
\begin{pmatrix}\cos\alpha & -\sin\alpha & 0 \\\sin\alpha & \cos\alpha & 0 \\0 & 0 & 1 \end{pmatrix}
\times \\
& & 
\times\begin{pmatrix}\cos\delta & 0 & -\sin\delta \\ 0 & 1 & 0 \\ \sin\delta & 0 & \cos\delta \\ \end{pmatrix}
\cdot
\begin{pmatrix}1 & 0 & 0 \\0 & \cos\rho & -\sin\rho \\0 & \sin\rho & \cos\rho\end{pmatrix},\nonumber
\end{eqnarray}
where the angles $\alpha$, $\delta$ and $\rho$ correspond to the 
right ascension, declination and field rotation. For simplicity, let us denote 
the field centroid coordinates (as obtained from the J2000 astrometric solution,
see the details above) with the same symbols. Let us denote the local
sidereal time (derived from UT1) by $\vartheta$. Then, it can be shown that 
the matrix
\begin{equation}
\mathbf{O}=\mathbf{A}(\alpha,\delta,\rho)^{\rm T}\mathbf{A}(\vartheta,\varphi_0,0)
\end{equation}
would be close to unity and the vector invariant of its logarithm
(i.e. the inverse of Eq.~\ref{eq:omatrixexp}) tells us the respective 
roll, pitch and yaw offsets that needed to be commanded to the hexapod. 
In practice, this vector invariant can be computed as follows. Let us 
define
\begin{eqnarray}
x & = & O_{32}-O_{23}, \\
y & = & O_{13}-O_{31}, \\
z & = & O_{21}-O_{12}, \\
r & = & \sqrt{x^2+y^2+z^2}, \text{and} \\
t & = & {\rm Tr}(\mathbf{O}) = O_{11} + O_{22} + O_{33}.
\end{eqnarray}
Then the components of the logarithm of $\mathbf{O}$ are going to be either 
\begin{equation}
\begin{pmatrix}P\\\Pi\\\Omega\end{pmatrix}=
\frac{{\rm arg}\left(t-1,r\right)}{r}
\begin{pmatrix}x\\y\\z\end{pmatrix}
\end{equation}
or identically zero when $x=y=z=0$. 

\subsection{Field tests}
\label{sec:tests}

In order to test both the constructed 
hexapod as well as the previously described 
algorithm, we attain a series of tests with a single camera mounted on the
hexapod payload platform (see also the right panel of Fig.~\ref{fig:hexapods}
for the actual setup).

During the determination of the numerical derivatives needed 
by Eq.~\ref{eq:omegaequation}, we mounted an $f/1.8$, $f=85\,{\rm mm}$
lens to a small-format camera having a pixel resolution of $4{\rm k}\times 4{\rm k}$
and a pixel size of $9\,{\rm \mu m}\times 9\,{\rm \mu m}$. This is an 
equivalent setup to the final construction of the Fly's Eye mosaic system 
with the exception that Fly's Eye camera units are using an $f/1.2$ lens 
instead of $f/1.8$. During these test runs, the lens were focused using the
spare test ports and pins found on the hexapod control electronics. 

Four image series have been taken according to the one-sided numerical
derivative needed by the approximation in Eq.~(\ref{eq:diffapproxlefthanded}).
All of the image series contain four individual frames acquired with
$20$ seconds of exposure time. In total, this procedure required $4\times 4$
frames while the gross time was roughly 10 minutes in total. However, 
during these experiments, the overall duty cycle was not ideal 
(for instance, the re-positioning of the hexapod between each series 
was not performed in parallel with the image readout and so). 
Based on these series of images we were able to obtain the field drift
and field rotation parameters $\omega_m$ and $\omega_m^{(A_k)}$. These
parameters were derived using the actual frame data (exposure durations
and observation time instances as found in the respective FITS header keywords) 
as well as the differential astrometric solutions obtained with the aid 
of the FITSH package \citep{pal2012}. Once we derived the $A_k$ coefficients,
we used these leg speeds (defined by Eq.~\ref{eq:speedperturbed})
to program the hexapod motion and obtain images with significantly longer
exposure times comparable to the proposed cadence of the Fly's Eye device.
One of the small stamps and the corresponding point-spread function (PSF)
is depicted in Fig.~\ref{fig:stamps}. One can clearly see that the 
procedure works smoothly and it is adequate for the intended goals
of the Fly's Eye all-sky survey. We repeated this experiment several times
by deliberately stirring the hexapod as well as we conducted similar tests
on subsequent nights.

In order to better characterize the quality of the sidereal tracking,
we performed an experiment by replacing the $f/1.8$, $f=85\,{\rm mm}$ 
lens by an $f/8$, $f=800\,{\rm mm}$ catadioptric optics. The usage of the 
latter type of lens yields a resolution of $2.3^{\prime\prime}$/pixel, i.e. 
roughly $9.4$ times finer than the intended Fly's Eye plate scale. 
Once the $A_k$ coefficients were properly derived using the $f=85\,{\rm mm}$
setup, the lens were replaced and a $\sim 3$ minute long sequence were taken
with exposure times of $30$ seconds. Since this setup yields a smaller 
field-of-view, we obtain only the field centroid drifts. Our tests show
that even with this finer resolution, the root-mean-square (RMS) deviation
of the frame sequences were always less than $0.3$ pixels, 
i.e. $0.7^{\prime\prime}$. 

According to the data acquisition strategy of the Fly's Eye device
\citep{pal2013}, frames are going to be synchronized to Greenwich sidereal time.
Hence, frames in every approx. $23$ hours and $56$ minutes should show exactly 
the same field (neglecting the effect of precession and nutation). 
Since the hexapod would make several hundreds of individual movements 
(including tracking, re-positioning and homing) between two frames 
with a cadence of exactly one sidereal day, even a few days long such 
series will provide us information about the longer term repeatability 
of the hexapod configuration. During our initial test runs, we performed
this analysis as well. We found that the RMS deviation between these 
corresponding frames is around $0.10-0.12$ pixels for the $f=85\,{\rm mm}$
lens. This deviation was obtained for a week long of consecutive run (i.e. 
the data acquisition was not interrupted by bad weather, but the system was 
shut down during daytime). While the primary purposes of this hexapod mount
and the classical telescope mounts are different, we can conclude that 
this RMS of $\sim2^{\prime\prime}$ is well comparable to large, meter-class
telescope pointing residuals \citep[see e.g.][and the references therein]{pal2015}. 
We note here that during this one-week run, the re-calibration procedure
(described in Sec.~\ref{sec:selfcalibration}) was not performed at all,
i.e. the same calibration constants $A_1$, $A_2$ and $A_3$ were involved
while commanding the hexapod. Therefore, we can expect that more
frequent attitude re-calibration decrease this residual, even in the
order of fractions of arcseconds (see in the following).

\subsection{Laboratory tests}
\label{sec:labtests}

Further laboratory tests were also performed using a very high resolution 
tiltmeter (HRTM) manufactured by Lippmann Geophysikalische Messger\"ate (L-GM). 
Such sensors are used in geodetics, more specifically, in astro-geodetic 
measurements \citep[see e.g.][]{hirt2008}. These sensors provide a white
noise output for pitch and roll tilts in the range of 
$0.01-0.02^{\prime\prime}/\sqrt{\rm Hz}$ and can even be 
used\footnote{beyond their intended purposes, such as levelling, seismology 
or the measurement of lunisolar tides} for metrology, i.e. quantify 
and validate devices capable of alter attitude. We performed repeatability 
tests with our hexapod when such a HRTM was mounted on the payload platform. 
Based on these quick tests, we conclude that the resolution in the
commanded attitude of this hexapod is in the range of 
$\lesssim 0.1^{\prime\prime}$ and the long-term
variations due to the ambient temperature changes are in the range of
$\sim 1^{\prime\prime}/^\circ{\rm C}$. 
Due to the parallel nature of the hexapod and the fact that 
actuation on all of the legs have an effect on the roll and pitch angles
(see the first two rows in the matrix $\mathbf{L}$ in Eq.~\ref{eq:l0matrix}),
such tests can be used to characterize the behaviour of each leg nearly
independently -- even if only 2 out of the 6 DoFs of the hexapod can 
be measured directly by such a tiltmeter device. However, one should be
aware that temperature \emph{inhomogeneties} in the device itself
could play an even more significant role in the precise response of the 
device to temperature changes. In order to accurately characterize this
dependence, we built and mount thermometers in all of the leg control
electronics (see Fig.~\ref{fig:blockdiagram}, where these thermometers
are connected to the ${\rm I^2C}$ bus) as well as in many points of the 
hexapod payload platform (including the 19 camera units in the final design). 
We expect that the frequent re-calibration (see Sec.~\ref{sec:absoltue}) in 
parallel with thermal data acquisition will provide sufficient information to
perform sub-arcsecond accuracy during routinely observations. 

All in all, we can conclude that our hexapod-based
sky tracking system is safely capable of providing sub-arcsecond tracking
precision. In addition, the hexapod based on this construction is capable
of performing measurements at the precision level needed by astro-geodetics
\citep[see also][for more details about such applications]{hirt2014}. 

\section{Summary}
\label{sec:summary}

This paper described how a hexapod can be used as a part of an all-sky
survey instrument to perform sidereal tracking. We show how can the 
astrometric information provided by the images be exploited in order 
to attain sub-arcsecond tracking precision. As a hexapod provides
a complete, 6 degrees-of-freedom motion control, the a survey instrument
based on this type of parallel kinematics can be used without any modifications
at arbitrary geographical location and without the need of precise alignment. 
We also demonstrated the mathematical background of the fault tolerant 
capabilities of a hexapod-based astronomical telescope mount. These 
capabilities are among the major advantages of using a hexapod 
in a fully autonomously and remotely operated astronomical instrument
since the sky tracking can continuously be performed even if one, two
or three of the hexapod legs are stuck, most likely because of
an electronic or mechanical failure.
Our mount design presented here forms the core part of the Fly's Eye project,
an ongoing moderate resolution all-sky time-domain variability survey. 

The benefits of a hexapod with respect to a conventional 
telescope mount are the following. First, conventional mounts are needed
to be extremely precisely aligned to the cardinal directions -- and if not,
even in the framework of an isotropic pointing model 
\citep{spillar1993,buie2003,pal2015}, the required corrections have a 
complexity which is comparable to that of a hexapod. In addition,
either an imprecisely aligned  equatorial mount or an alt-az mount 
require a field rotation mechanism for compensating these errors. While
in the case of a small field-of-view telescope, this third stage of serial 
kinematics providing the field rotation compensation 
is significantly smaller than the drivers of the two main axes, it is not
true for an all-sky instrument. In other words, off-the-shelf field rotators
cannot be applied at all for an all-sky device. 
Another benefit of using hexapods instead
of conventional mounts is the fact that the characteristic alignment of the
payload platform is always horizontal in its home position. In the case 
of an equatorial mount it is not true and even this direction
depends on the actual geographical location of the installation. Therefore,
the same design cannot be used at different locations if the payload is
a camera assembly performing all-sky surveying. In the case of alt-az mounts,
the preferred direction of an all-sky camera (i.e. looking towards the zenith)
coincides with the gimbal lock position of the telescope mount (even 
considering a field rotation mechanism), hence alt-az mounts are intrinsically
not capable of performing such a tracking. Of course, while the issues 
mentioned above can be overcome \citep[see e.g.][where a conventional 
equatorial mount is used to support a mosaic camera system]{law2015}, 
hexapods can provide a cost-effective and elegant 
way of long-term maintenance of all-sky instrumentation which also feature
precise sidereal tracking. 

\acknowledgments
We thank Hans Deeg and \v{Z}eljko Ivezi\'c for the useful discussions
regarding to the design itself and the practical scientific implications 
of the Fly's Eye project. Our research is supported mainly by the 
Hungarian Academy of Sciences via the grant LP2012-31. Additional support
is also received via the OTKA grants K-109276, K-104607 and K-113117. 
In our project, we involved numerous free \& open source software, including
gEDA (for schematics and PCB design), FreeCAD (3D designs), 
CURA (3D slicing, GCODE generation and printing control) and 
AVR-GCC (for MCU programming).

{}



\begin{thebibliography}{}
\bibitem[Bieryla et al.(2015)]{bieryla2015}
Bieryla, A. et al.
2015, \aj, 150, 12

\bibitem[Buie(2003)]{buie2003}
Buie, M.: \emph{General Analytical Telescope Pointing Model}, available from
http://www.boulder.swri.edu/\~{ }buie/idl/
/downloads/pointing/pointing.pdf

\bibitem[Chini(2000)]{chini2000}
Chini, R. 
2000, \rma, 13, 257

\bibitem[Deeg et al.(2004)]{deeg2004}
Deeg, H. J.; Alonso, R.; Belmonte, J. A.; Alsubai, K.; Horne, K. and Doyle, L.
2004, \pasp, 116, 985

\bibitem[Geijo et al.(2006)]{geijo2006}
Geijo, E. M. et al.
2006, \procspie, 6273, 38

\bibitem[Granzer et al.(2012)]{granzer2012}
Granzer, T. et al.
2012, Astron. Nachtr., 333, 823

\bibitem[Hirt \& Seeber(2008)]{hirt2008}
Hirt, Ch. \& Seeber, G. 
2008, J. Geodesy, 82, 347

\bibitem[Hirt et al.(2014)]{hirt2014}
Hirt, Ch.; Papp, G.; P\'al, A.; Benedek, J. \& Sz\H{u}cs, E.
2014, Meas. Sci. Technol., 25, id. 085004

\bibitem[Huang, Raffin \& Chen(2011)]{huang2011}
Huang, Y.-D.; Raffin, Ph. \& Chen, M.-T.
2011, IEEE Transactions on Antennas and Propagation, vol. 59, issue 6, pp. 2022-2028

\bibitem[Ivezi\'c et al.(2008)]{ivezic2008}
Ivezi\'c, \v{Z}. et al.
2008, LSST: from Science Drivers to Reference Design and Anticipated Data Products, arXiv:0805.2366

\bibitem[Koch et al.(2009)]{koch2009}
Koch, P. M. et al. 
2009, \apj, 694, 1670

\bibitem[Kaiser et al.(2002)]{kaiser2002}
Kaiser, N. et al.
2002, \procspie, 4836, 154

\bibitem[Law et al.(2012)]{law2012}
Law, N. M. et al.
2012, Ground-based and Airborne Telescopes IV. Proceedings of the SPIE, Volume 8444, article id. 84445C

\bibitem[Law et al.(2015)]{law2015}
Law, N. M. et al.
2015, PASP, 127, 234

\bibitem[\protect\citeauthoryear{Meeus}{1998}]{meeus1998}
Meeus, J.:
\emph{Astronomical algorithms} (2nd ed.), 1998, Richmond, VA: Willmann-Bell.


\bibitem[M\'esz\'aros et al.(2014)]{meszaros2014}
M\'esz\'aros, L.; Jask\'o, A.; P\'al, A. \& Cs\'ep\'any, G.
2014, \pasp, 126, 769

\bibitem[P\'al(2012)]{pal2012}
P\'al, A. 
2012, \mnras, 421, 1825

\bibitem[P\'al et al.(2013)]{pal2013}
P\'al, A. et al.
2013, \an, 334, 932

\bibitem[P\'al et al.(2015)]{pal2015}
P\'al, A., Vida, K., M\'esz\'aros, L. \& Mez\H{o}, Gy.
2015, Exp. Astron., in press (arXiv:1507.05469)

\bibitem[Pepper et al.(2007)]{pepper2007}
Pepper, J. et al.
2007, \pasp, 119, 923

\bibitem[Spillar et al.(1993)]{spillar1993}
Spillar, E. J. et al.
1993, \pasp, 105, 616

\bibitem[Vida et al.(2014)]{vida2014}
Vida, K. et al.
2014, Proceedings of ``Observing techniques, instrumentation and science for metre-class telescopes'',
Contributions of the Astronomical Observatory Skalnat\'e Pleso, 43, 530

\bibitem[\protect\citeauthoryear{Wallace}{2008}]{wallace2008}
Wallace, P. T.
2008, Advanced Software and Control for Astronomy II. Edited by Bridger, Alan; Radziwill, Nicole M. Proceedings of the SPIE, Volume 7019, article id. 701908, 12 pp.

\end{thebibliography}
\end{document}